  \documentclass[12pt,showpacs,preprintnumbers,amsmath,amssymb]{revtex4}
 \usepackage{graphicx}

\begin{document}

\title{Model of e-cloud instability in the Fermilab Recycler }
\author{V.~Balbekov}
\affiliation {Fermi National Accelerator Laboratory\\
P.O. Box 500, Batavia, Illinois 60510}
\email{balbekov@fnal.gov} 
\date{\today}
\begin{abstract}
\begin{center}
{\bf Abstract}
\end{center}

\noindent
Simple model of electron cloud is developed in the paper to explain e-cloud 
instability of bunched proton beam in the Fermilab Recycler \cite{MAIN}. 
The cloud is presented as an immobile snake in strong vertical magnetic field.
The instability is treated as an amplification of the bunch injection errors 
from the batch head to its tail.
Nonlinearity of the e-cloud field is taken into account.
Results of calculations are compared with experimental data demonstrating 
good correlation.

\end{abstract}
\maketitle

%

\section{Introduction}

%
Fast transverse instability of proton beam in the Fermilab Recycler 
has been observed and reported recently \cite{MAIN}.
Convincing arguments are adduced that the instability is caused by electron clouds.
However, a detailed theoretical analysis is not performed in the quoted work.
Development of theoretical model which is capable to explain these 
data in a consistent way is the aim of this note.  
 
%

\section{Electron cloud model}

%
Transverse cross sections of e-cloud in the Recycler are represented in Fig.~A1
of the Appendix.
The figures are copied from Ref.~\cite{MAIN} where they have been obtained 
by computer simulation with POSINST code \cite{POS}.

The main conclusion follows from these pictures that the e-cloud 
density almost does not depend on vertical coordinate $\,y$,
especially inside the proton beam which is sketched as the red circle.
This feature of e-cloud in strong magnetic field is confirmed in other 
papers (see e.g. \cite{POL}) and is in a consent with following simple explanation.

Transverse motion of electrons inside a proton bunch 
in the presence of vertical magnetic field $\,B\,$ is described by the equations
%
\begin{equation}
\ddot x_e+\omega_B^2 x_e=-\omega_e^2(x_e-X_p),\qquad
 \ddot y_e=-\omega_e^2(y_e-Y_p)
\end{equation} 
%
with the coefficients
%
\begin{equation}
 \omega_B^2 = \left(\frac{eB}{m_ec}\right)^2,\qquad 
 \omega_e^2 = \frac{2\pi e^2\rho_p}{m_e}
\end{equation} 
%
The simplest model of proton beam as a rod of constant density $\,\rho_p\,$
centered in the point $\,(X_p,Y_p)\,$ is used here.
However, it will be shown soon that it is an assumption of a little importance.

The proton beam in the Recycler can oscillate with betatron frequency $\,\omega_p$. 
Therefore relations of amplitudes following from Eq.~(1) are 
%
\begin{equation}
 \frac{x_e}{X_p}=\frac{\omega_e^2}{\omega_B^2+\omega_e^2-\omega_p^2},\qquad
 \frac{y_e}{Y_p}=\frac{\omega_e^2}{\omega_e^2-\omega_p^2}
\end{equation} 
%
The parameters taken for the following numerical estimations are:
$\,B=0.145\,{\rm T},\;\rho_p=1.3\times 10^{15}/{\rm m}^3$.
Angular velocity of protons in the Recycler can be used as a convenient unit:
$\Omega=2\pi\times 89.8\,{\rm kGz} = 0.564\times 10^6/{\rm s}$. 
Then $\;\omega_B=\Omega Q_B,\;\omega_e=\Omega Q_e\,$ where 
$\,Q_B\simeq 45000,\;Q_e\simeq 2500$.
Because the proton beam tune is $\,Q\simeq 24.4$, the amplitude ratios are:
%
\begin{equation}
   \frac{x_e}{X_p}=\frac{Q_e^2}{Q_B^2+Q_e^2-Q^2}\simeq 0.003, \qquad 
   \frac{x_e}{X_p}=\frac{Q_e^2}{Q_e^2-Q^2}\simeq 1.0001  
\end{equation}
%
It means that the movement of electrons is awfully obstructed in horizontal 
direction due to magnetic field.
As for vertical direction, each electron follows the protons bunch 
when it is located inside it, and moves ``free'' between the bunches.
In such a manner it can reach the pipe walls where it can drive out secondary 
electrons which can be trapped by next bunch, etc.  
As a result, each primary electron creates a vertical e-stream 
composed of secondary electrons.
The stream density depends on time but it keeps a fixed position in $(x$-$z)$ 
plane coinciding with position of the proton which was begetting the primary electron.
A host of the streams forms a stationary e-cloud which transverse cross-section 
is presented in Fig.~A1.

The cloud top view is sketched in Fig.~1 where several proton bunches are shown, 
each oscillating horizontally due to injection errors.
The bunches create snake-like immovable e-traces as they are drafted in the picture.
The traces of several bunches coincide if they have been injected with the 
same errors, or differ from each other if the errors differ
(bunch \#2 in Fig.~1).
According to the model, electron density at distance $\,s\,$ from the cloud 
beginning can be represented in the form
%
\begin{figure}[t!]
\includegraphics[width=100mm]{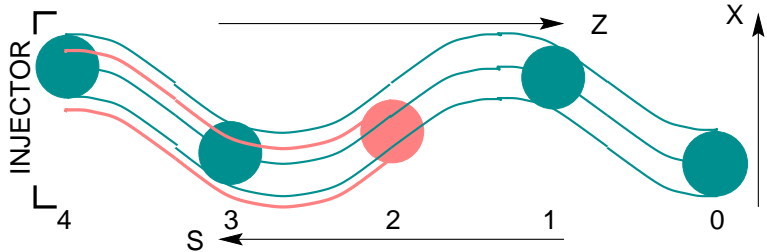}
\caption{Top view of e-cloud. Each proton bunch in the picture generates an 
immovable e-snake.
The snakes coincide each other if their parent bunches have the same injection 
conditions, and differ otherwise~(\#2). 
Local density of each snake depends on time.}
\end{figure}
%
%
\begin{equation}
 \rho_e(x,s,t) = e\int_0^s 
 w(\tau)\,\bar\rho\big(x-X(s,t-\tau)\big)\lambda(s')\,ds',\qquad
 \tau=\frac{s-s'}{v}
\end{equation}
%
where $\,\bar\rho(x)\,$ is steady state (w/o coherent oscillations) projection 
of proton beam on axis $\,x$, $\,X(s,t)\,$ is the beam coherent displacement,
and $\,\lambda(s)\,$ is its linear density. 
The coefficient $\,w(\tau)\,$ describes evolution of each snake local density 
which appears, increases in a time due to secondary electrons, and decays 
eventually because the e-oscillations are typically unstable as they are focused
by field of bunched proton beam.  
Calculation of this function is not a subject of this paper, and it will be 
treated further as some phenomenological parameters.

%

\section{Proton equations of motion}

%
Horizontal electric field of the cloud looks like Eq.~(5)
%
\begin{equation}
 E_e(x,s,t) = e\int_0^s w(\tau)\,F\big(x-X(s,t-\tau)\big)\lambda(s')\,ds',
 \qquad F'(x)=4\pi\bar\rho(x)
\end{equation}
%
If the beam consists of short identical bunches, the integral turns into the sum
%
\begin{equation}
 E_N(t,x) = e\sum_{n=0}^{N}w_n F\Big(x-X_{N-n}(t-nT)\Big)
\end{equation}
%
where $T$ is the time separation of the bunches which are enumerated from 
the beam head (index 0) to the current bunch (index $N$).
Therefore equation of horizontal oscillations of a proton in $\,N^{\rm th}$ 
bunch is  
%
\begin{equation}
 \ddot x(t)+\omega_0^2x =
-\frac{e^2}{m\gamma}\sum_{n=0}^{N}w_n F\Big(x-X_{N-n}(t-nT)\Big)
\end{equation}
%
where $\omega_0$ is betatron frequency without e-cloud.
It gives for small oscillations:
%
\begin{equation}
 \ddot x(t)+\omega_0^2x=-2\omega_0\sum_{n=0}^{N} W_n\Big[x-X_{N-n}(t-nT)\Big]
\end{equation}
%
where $W_n=4\pi e^2\rho_n(0)w_n/m\gamma$, and $\rho_n(x)$ is averaged over $s$ 
e-cloud density in $n^{\rm th}$ bunch.
Therefore average incoherent betatron frequency is in $N^{\rm th}$ bunch:
%
\begin{equation}
 \omega_N = \Big(\omega_0^2+2\omega_0\sum_{n=0}^{N} W_n\Big)^{1/2}
 \simeq \omega_0+\Delta\omega_N, \qquad \Delta\omega_N=\sum_{n=0}^{N} W_n
\end{equation}
%
Thus the coefficient $\,W_n\,$ has to be treated as the betatron frequency shift 
created by a bunch \#$(N-n)$ in the bunch \#$N$.
It is clear that this single-bunch wake has a restricted length which effectively
can be taken as $\,N_w$.    
Then $N_w\overline W\,$ is the saturated tune shift which could be created by rather
long batch and measured experimentally ($\overline W\,$ is the average value). 

%

\section{Coherent oscillations (linear approximation)}

%

Small betatron oscillations are considered in this section.
Averaging Eq.~(9) over each bunch, one can obtain equations of motion 
of the bunch centers:
%
\begin{equation}
 \ddot X_N+\omega_0^2X_N =
-2\omega_0\sum_{n=0}^{N} W_n\big[X_N(t)-X_{N-n}(t-nT)\big],
 \qquad N=0,\,1,\,\dots
\end{equation}
%
Looking general solution in the form
%
\begin{eqnarray}
 X_N(t)=A_N(t) \exp\big(i\omega_0[t-nT]\big)+{\rm c.c.} 
\end{eqnarray}
%
and using ordinary method of averaging,
one can get series of equations for the complex amplitudes:
%
\begin{eqnarray}
 \dot A_N(t)=i\sum_{n=0}^{N}W_n\big[\,A_N(t)-A_{N-n}(t-nT)\big]
\end{eqnarray}
%
The special solution of the series has to be emphasized particularly: $A_N(t)$~=~const
that is the amplitudes depend neither time nor bunch number.
It means that all bunches are injected at the same conditions and move 
on the same trajectory, as it is shown in Fig.~1 by solid lines.
Field on axis of this e-cloud is zero so it does not affect motion of the bunch 
centers.
It follows from this that some spread of the injection conditions is one of the 
prerequisites of the e-cloud instability. 
 
It should be emphasized in this connection that coherent eigentunes of the 
bunches coincide with their incoherent tunes (if the bunch coherent interaction 
is excluded).
It is apparent that the mutual influence of bunches is stronger when their 
eigentunes are closer.
Therefore an approach of the eigentunes is another prerequisite of the 
instability. 

Fig.~2 is represented for an explanation of these statements.
A possible wake function of a single bunch is sketched separately in upper part 
of the picture.
It arises inside the bunch, remains constant for a time, and quickly decays 
after this. 
The figure itself represents corresponding e-cloud of a long batch. 
The density (and the tune shifts) increase in the beginning of the batch
(5 bunches in the example), and remains constant hereafter. 
%
\begin{figure}[b!]
\includegraphics[width=140mm]{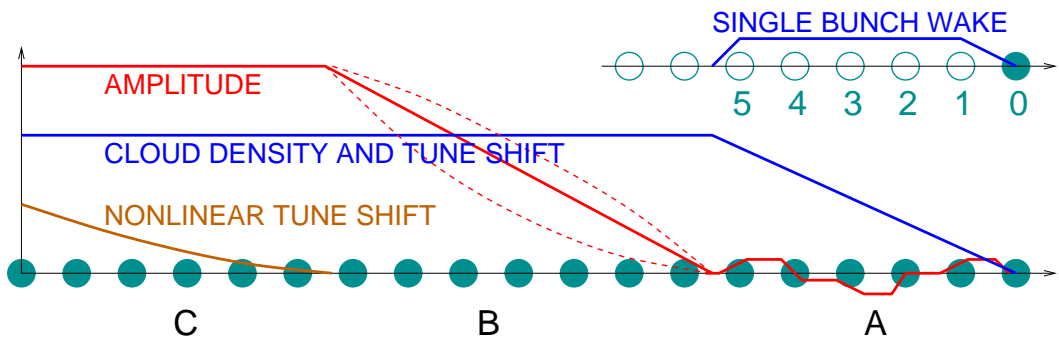}
\caption{Proton batch and its e-cloud (very schematically). 
Top -- single bunch wake: rise, growth, being, decay (lifetime 5 periods).
Bottom blue line -- e-cloud density and tune shift: 
(A) 5 period growth, (B+C) saturation.
Red line -- amplitude of coherent oscillations: (A) irregular wobbling at different 
tunes of the bunches; (B) more or less systematic growth at coinciding tunes;
(C) saturation due to nonlinearity of the wake field.}
\end{figure}
%
The red line is added to sketch expected behavior of coherent amplitudes. 
Rather disorderly moving in the beginning of the batch changes into a systematic 
growth later.
It is taken into account also that the growth cannot be unrestricted, in particular
because of nonlinear effects which are beyond of presented approximation 
and will be considered in following section.

This pattern has to be compared with results of e-cloud simulation presented 
in Fig.~A2.
Even though the simulated shape of the e-cloud is more complicated, the model 
reflects its important properties: fast growth in the beginning and saturation 
at the end (partial decay and oscillations between the bunches seem to be 
less important for the bunch centers).

%

\subsection{Constant wake.}

%
\begin{figure}[b!]
\includegraphics[width=100mm]{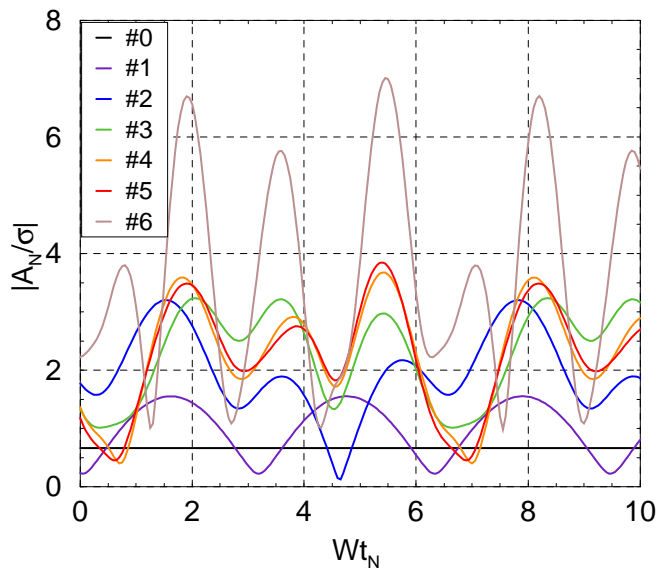}
\vspace{-7mm}
\caption{Amplitudes of bunch oscillations with constant wake. 
Initial amplitudes have Gaussian distribution with dispersion $\,\sigma\,$ 
(both real and imaginary parts).}
\end{figure}
%
The case $\,W_n=W=const\,$ is considered in this subsection as a preliminary step.
It means that the wake of any bunch does not decay, at least in the range 
of considered part of the batch.
Then series (13) obtains the form:
%
\begin{eqnarray}
 \dot A_N(t)=iW\sum_{n=0}^{N}\big[\,A_N(t)-A_{N-n}(t-nT)\big]
\end{eqnarray}
%
and has the simple solution
%
\begin{equation}
 A_N(t) = A_{0i}+\sum_{n=1}^N\;(A_{ni}-A_{n-1,i})\exp(inWt_N)
\end{equation}
%
where $\,t_N=t-NT$, and subindex 'i' marks the injection instance $\,t=nT\,$
that is $\,t_n=0$. 
This result affirms both of foregoing statements: (i) e-cloud does not affect 
coherent oscillation at stable injection conditions, (ii) systematic growth 
of the coherent amplitudes does not happen at different eigentunes of the bunches.
An example is given in Fig.~3 where random distribution of initial amplitudes has 
been applied.

%

\subsection{One-step wake}

%
An opposite case is considered in this subsection: very short wake which can reach
only the nearest following bunch: $\,W_n=W\delta_{n,1}$. 
Then Eq.~(13) gives:
%
\begin{eqnarray}
 \dot A_0=0,\qquad\dot A_{N>0}(t)=iW\,\big[\,A_N(t)-A_{N-1}(t-T)\,\big]
\end{eqnarray}
%
which series has the solution
%
\begin{eqnarray}
 A_0=A_{0i},\qquad        A_{N}(t)=A_{0i}
+\exp\big(iWt_N)\sum_{n=1}^N\frac{A_{ni}-A_{0i}}{(N-n)!}\,(-iWt_N)^{N-n}  
\end{eqnarray}
%
It is seen that  systematic amplitude growth is possible at $N\ge2$ with additional 
condition that initial complex amplitude of the bunch, or at least one of 
the previous bunches, differs from amplitude of the leading bunch $\,(N=0)$. 
Examples are given in Fig.~4 for the conditions $\,A_{ni}=1-\delta_{n0}\,$
that is the injection error is 0 at leading bunch, and 1 at others.
Another example is given by Fig.~5 with random initial distribution.
It looks very similarly in average though a random spread appears.
%
\begin{figure}[t!]
\includegraphics[width=100mm]{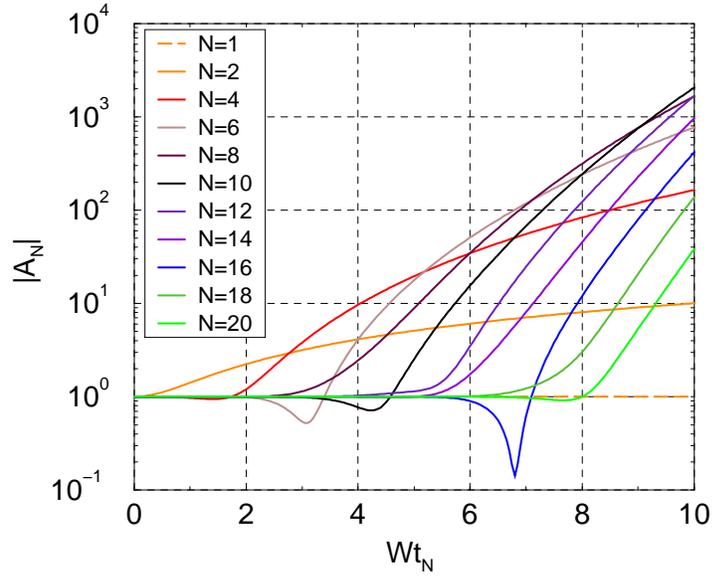}
\vspace{-5mm}
\caption{Amplitudes of bunch oscillations at initial values 
$A_{Ni}=\delta_{N0}$. The leading bunch $(N=0)$ does not oscillate having zero
error of injection, but it creates a path for following bunches 
providing them the same eigentune. 
Coherent betatron amplitude is constant at first bunch $(N=1)$, 
growths linearly at $N=2$, etc.}
\end{figure}
%
\begin{figure}[h!]
\includegraphics[width=100mm]{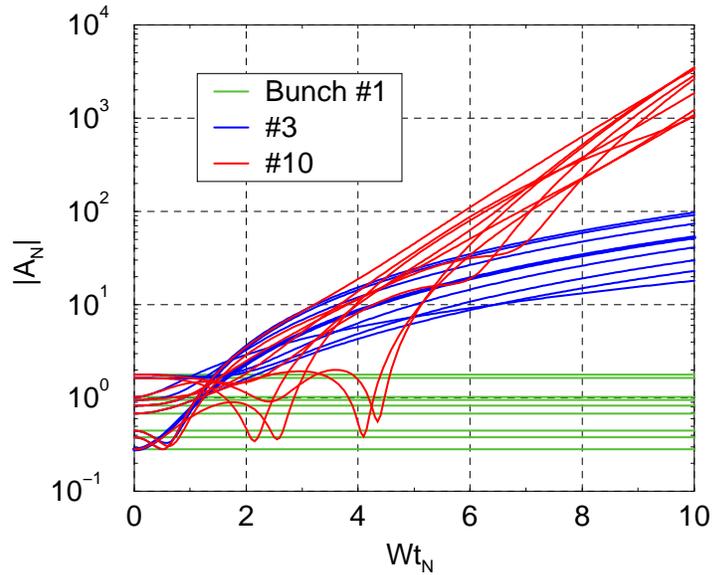}
\vspace{-5mm}
\caption{Amplitudes of bunch oscillations with a random errors of injection.
Initial amplitude of the leading bunch $A_{0i}=0$, other initial amplitudes
are distributed randomly in the interval $\,-1<A_{Ni}<1$. 
Bunches $N=1,$~3, and 10 are presented for 10 random realization.}
\end{figure}
%
%

\subsection{Restricted uniform wake}

%
\noindent
Restricted uniform wake is considered in this subsection as the more general case:
%
\begin{equation}
 W_n = \bigg\{{W \quad{\rm at}\quad n\le N_w
       \atop  \;0\quad\;\;{\rm at}\quad n>N_w   }
\end{equation}
%
Then Eq.~(13) obtains the form
%
\begin{eqnarray}
 \dot A_N(t)=iW\sum_{n=0}^{N_{m}}\big[\,A_N(t)-A_{N-n}(t-nT)\big],
 \qquad N_{m} = {\rm min}\{N,N_w\}
\end{eqnarray}

%

\subsubsection{The batch head}
\noindent
%
At $\,N\le N_w,$ Eq.~(19) coincides with Eq.~(14) having solution (15).
These amplitudes can oscillate only at variable injection conditions (Fig.~3). 
%

\subsubsection{The batch tail.}
\noindent
%
Thus there is no systematic growth of the amplitude in the butch head, 
but there is only same variation due to random superpositions of 
the bunch amplitudes.
Obtaining amplitudes are as small as injection errors, and can be 
neglected at the analysis of the batch tail.
Then it is convenient to use new bunch enumeration $N'=N-N_w=1,\,2,\,\dots\,$, 
and modified complex amplitudes
%
\begin{eqnarray}
 A'_{N'}(t) = A_{N'+N_w}(t)\exp\big(-iN_wW[t-NT]\big),\qquad N'=1,\,2,\,\,...
\end{eqnarray}
%
These variables satisfy the series of equations: 
%
\begin{eqnarray}
 \dot A'_{0}(t)=0,\qquad\dot A'_{N'}(t)=-iW\sum_{n=1}^{N_m}A'_{N'-n}(t-nT),
 \qquad N_m={\rm min}\{N_w,N'\}
\end{eqnarray}
%
The solutions can be represented in the form
%
\begin{equation}
 A'_{N'}(t)=\sum_{k=0}^{N'}\frac{S_{N'}^{(k)}}{k!}\big[(-iW(t-NT)\big]^k
\end{equation}
%
\begin{table}[b!]
\vspace{-5mm}
\begin{center}
\caption{Coefficients $S_N^{(k)}$ at initial amplitudes $a_{Ni}=1$.}
\vspace{5mm}
\begin{tabular}{|p{12mm}|c|c|c|c|c|c|c|c|c|c|}
\hline 
 $\;N'\rightarrow$   &~~~0~~~&~~~1~~~&~~~2~~~&~~~3~~~&~~~4~~~&~~~5~~~
                        &~~~6~~~&~~~7~~~&~~~8~~~&~~~9~~~            \\
\hline
 $\;k=0$            & 1 & 1 & 1 & 1 & 1 & 1 & 1 & 1 & 1 & 1         \\
\hline
 $\;k=1$            & - & 1 & 2 & 3 & 4 & 5 & 5 & 5 & 5 & 5         \\
\hline
 $\;k=2$            & - & - & 1 & 3 & 6 &10 &15 &19 &22 &24         \\
\hline
 $\;k=3$            & - & - & - & 1 & 4 &10 &20 &35 &53 &72         \\
\hline
 $\;k=4$            & - & - & - & - & 1 & 5 &15 &35 &70 &122        \\
\hline
 $\;k=5$            & - & - & - & - & - & 1 & 6 &21 &56 &126        \\
\hline
 $\;k=6$            & - & - & - & - & - & - & 1 & 7 &28 &84         \\
\hline
 $\;k=7$            & - & - & - & - & - & - & - & 1 & 8 &36         \\
\hline
 $\;k=8$            & - & - & - & - & - & - & - & - & 1 & 9         \\
\hline 
 $\;k=9$            & - & - & - & - & - & - & - & - & - & 1         \\
\hline
\end{tabular}
\end{center}
\end{table}
%
\begin{figure}[t!]
\includegraphics[width=100mm]{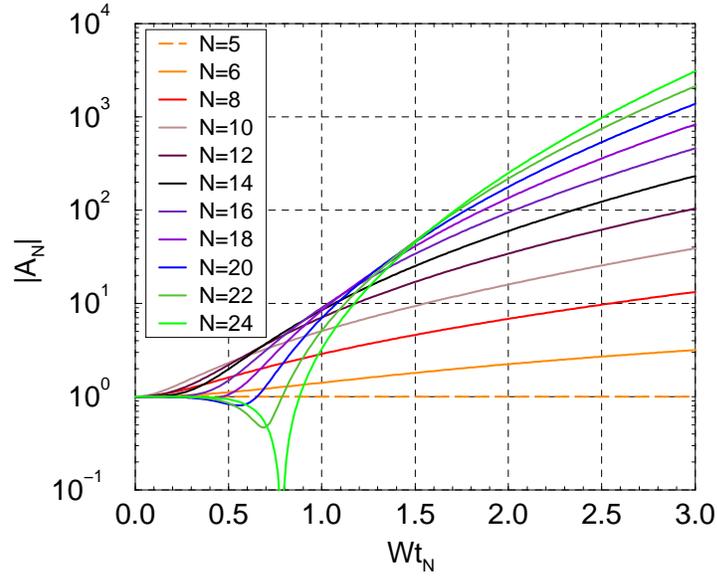}
\vspace{-5mm}
\caption{The tail part of the batch with 5-step wake. Initial amplitudes 
of presented bunches $A_{N'i}=1$, foregoing bunches do not oscillate.}
\end{figure}
%
\begin{figure}[h!]
\includegraphics[width=100mm]{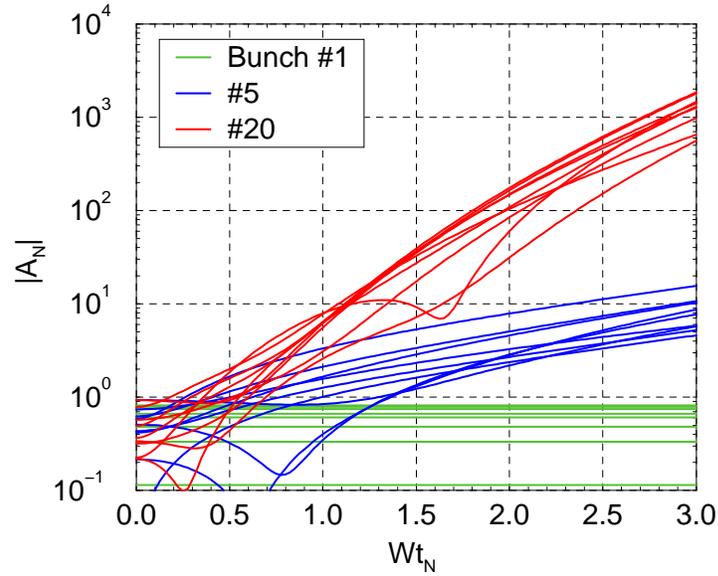}
\caption{The same conditions as in Fig.~6 but initial amplitudes
are random numbers distributed uniformly in the interval $\,-1<A_{N'i}<1$. 
Bunches $N=1,$~5, and 20 are presented for 10 random realization}
\end{figure}
%
$\!\!\!$with coefficients satisfying the series of equations
%
\begin{equation}
 S_{N'}^{(k)} = \sum_{n=1}^{N'_m}S_{N'-n}^{(k-1)},  
 \qquad N'_m={\rm min}\{N_w,N',N'+1-k\}
\end{equation}
%
Set of the coefficients $S_{N'}^{(0)}$ should be defined separately to satisfy 
initial conditions.
In accordance with Eq.~(20)~and~(22) 
%
\begin{equation}
 S_{N'}^{(0)}=A'_{N'}(NT)=A_{N'+N_w}(NT)=A_{Ni}
\end{equation}
%
where subindex $i$ marks initial amplitude of $N$-th bunch at $\;t=TN$.
Calculation of coefficients $\,S_{N'}{(k)}\,$ is not a problem 
with any initial conditions because Eq.~(23) is actually a recurrent relation. 
An example is given in Table I with the parameters: 
$\,N_w=5,\;\;A_{Ni}=0\,$ at $\,N<N_w\,$ or $\,A_{Ni}=1\,$ at $\,N\ge N_w$,
Corresponding dependence of amplitudes on time is plotted in Fig.~6. 
Another example is presented in Fig.~7 with a random initial distribution.
There is some resemblance of these plots to Fig.~4-5 where one-step wake has been
represented.
However, the amplitude growth rate is now about 3 times faster at the same $W$.

%
\vspace{-5mm}
\subsection{Discussion}

%

The main conclusions from this model are:
\\
1. The ``instability'' appears because of injection errors which are 
amplified from bunch to bunch along the batch.
\\
2. However, it cannot appear at absolutely stable injection conditions. 
Some spread of the errors is necessary to turn on the bunch coherent interaction.
\\
3. Expected bunch coupling is not very strong in the batch beginning because of
essential difference of their eigentunes.    
Non-growing bunch oscillations are possible in this part. 
The amplitudes coincide with injection errors in order of value,
having an interference if the errors are varied.
Duration of this part is about saturation time of the cloud.
\\
4. Systematic growth of the amplitudes can happen in more remote parts of the 
batch where the e-cloud is saturated.
The amplitude growth rate increases from the batch beginning to its tail
being almost exponential at the end. 

These statements are partially confirmed by experimental data presented 
in Fig.~A3.
Very small and about constant amplitudes are observed in the front part 
of the batch including about 20 bunches.
Then the amplitudes demonstrate a growth by factor 2-3 in following 20-30 
bunches.
These data do not conflict to the model with $\,N_w\simeq 20$. 
However, the bunches with $\,N>\sim 50\,$ have about equal amplitudes  
whereas the model predicts their unrestricted growth.
Similar contradiction (saturation against unrestricted growth) concerns the long 
term effects that is the dependence of the amplitudes on revolution number.
Nonlinearity of the e-cloud field is a possible cause of the contradictions,
which statement will be considered in following section.
 
%

\section{Nonlinear effects (1 step wake).} 

%
With cubic nonlinearity taken into account, Eq.~(8) gives the equation of
betatron oscillation of protons in $\,N^{\rm th}\,$ bunch:
%
\begin{equation}
 \ddot x(t)+\omega_0^2x(t)=-2\omega_0\sum_{n=0}^N W_n
 (\xi_n+\epsilon_n\xi_n^3/3)
\end{equation}
%
where
$ 
 \hspace{25mm} \xi_n = x(t)-X_{N-n}(t-nT),\qquad \epsilon_n=\rho_e"(0)/2\rho_e(0)
$
\\
Solution of this equation can be represented in the form like Eq.~(12):
%
\begin{equation}
 x(t) = a(t)\exp\big(i\omega_0[t-NT]\big)+c.c.
\end{equation}
%
providing the equation for amplitude $\,a$ 
%
\begin{equation}
 \dot a(t) = i\sum_{n=0}^N W_n\eta (1+\epsilon_n|\eta_n|^2),    
 \qquad \eta_n=a(t)-A_{N-n}(t-nT).
\end{equation}
%
One-step wake is considered below: $\,W_n=W\delta_{n1}$.
Note that the condition $\,W_0=0$ follows from this definition 
which is reasonable because a noticeable e-cloud cannot appear in 
the leading bunch in absence of secondary electrons.
Therefore amplitude of any particle does not depend on time in this bunch,
so that amplitude of the bunch center is constant as well.
The last can be taken as 0 because a difference of other bunches is the only crucial 
circumstance.
Their equations of motion are 
%
\begin{equation}
 \dot a(t) = iW\big[\,a(t)-A_{N-1}(t-T)\,\big]
 \big[\,1+\epsilon\big|a(t)-A_{N-1}(t-T)|^2\big],
 \qquad A_0=0.
\end{equation}
%
Following steps have been used for numerical solution of the equations:
\\
1. Generation of random initial distribution of particles in first bunch;
\\
2. Calculation of the function $\,a(t)\,$ for each particle of this bunch $(N=1)$
by solution of Eq.~(28) with the known $A_{N-1}=A_0$;
\\
3. Calculation of the central amplitude $\,A_1(t)\,$ as a function of time;  
\\
4. The same for second bunch with known $\,A_1(t)$, etc.
\\
Results of the calculation are presented below.
  
Instability of a thin beam $(R=0)$ is illustrated by Fig.~8 where the bunch offsets 
are taken as $\,A_N=1\,$, and the nonlinearity parameter $\,\epsilon=-0.001\,$.
The left-hand graph represents dependence of amplitudes of the bunches on time. 
It can see that their behavior at $\,Wt_N<\sim 4$ closely resembles the curves of
Fig.~4 where similar beam is considered without nonlinearity.
However, the plots strongly differ further because the growth of the amplitudes
in the nonlinear system actually deceases at $\,A\simeq30\,$ 
(note that corresponding to $\epsilon A^2\simeq-1$).

The amplitude averaged over all the bunches and its growth rate are shown in 
the right-hand picture. 
It is seen that the rate peaks at $\,Wt\simeq 4$, and it is about 0 at $\,Wt=10$. 

Motion of second bunch is considered more closely in Fig.~9
where its amplitude against time (left-hand graph) and phase trajectories
are presented at different nonlinearity.
It is a typical behavior of nonlinear oscillator exited by periodical external
field which cannot be treated as Landau damping. 

A thick beam is considered at the same conditions being presented by similar 
plots in Fig.~10. 
The water-bag model of radius 1 is used for transverse distribution of the 
proton beam.
There is no essential difference between Fig.~8~and~9, demonstrating that the beam 
radius is a factor of second importance in this problem.

Next example pertains to the same beam with different injection errors of the bunches: 
$\,A_{Ni}=0.3+0.1\exp(i\phi_N)\;$ with random phase $\,\phi_N$. 
One of the random realization is presented in Fig.~11 which has only small 
distinction from previous examples.
%
\begin{figure*}[t!]
\hspace{-10mm}
\begin{minipage}[b!]{0.45\linewidth}
\begin{center}
\includegraphics[width=85mm]{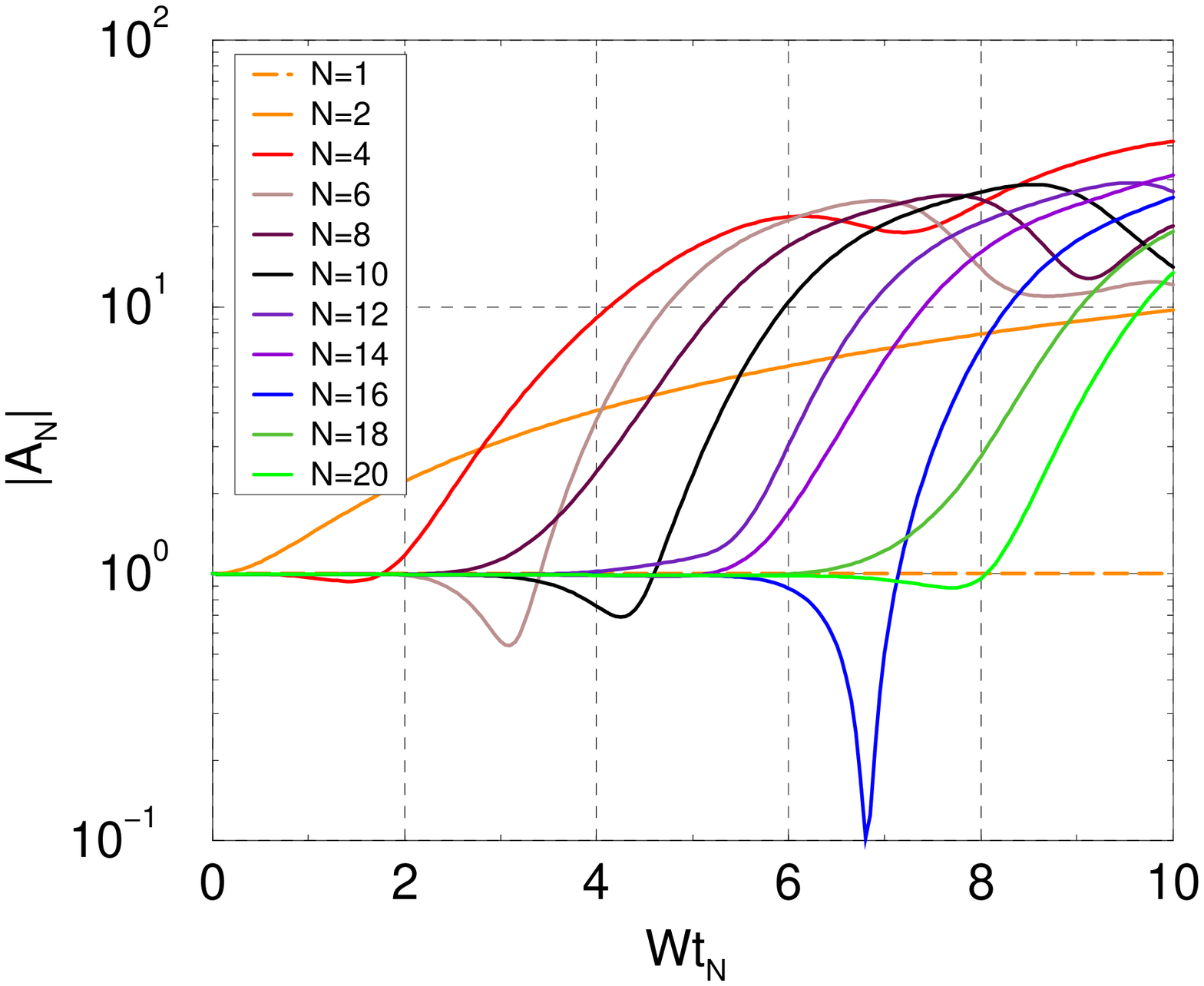}
\end{center}
\end{minipage}
\hspace{0mm}
\begin{minipage}[b!]{0.45\linewidth}
\begin{center}
\includegraphics[width=85mm]{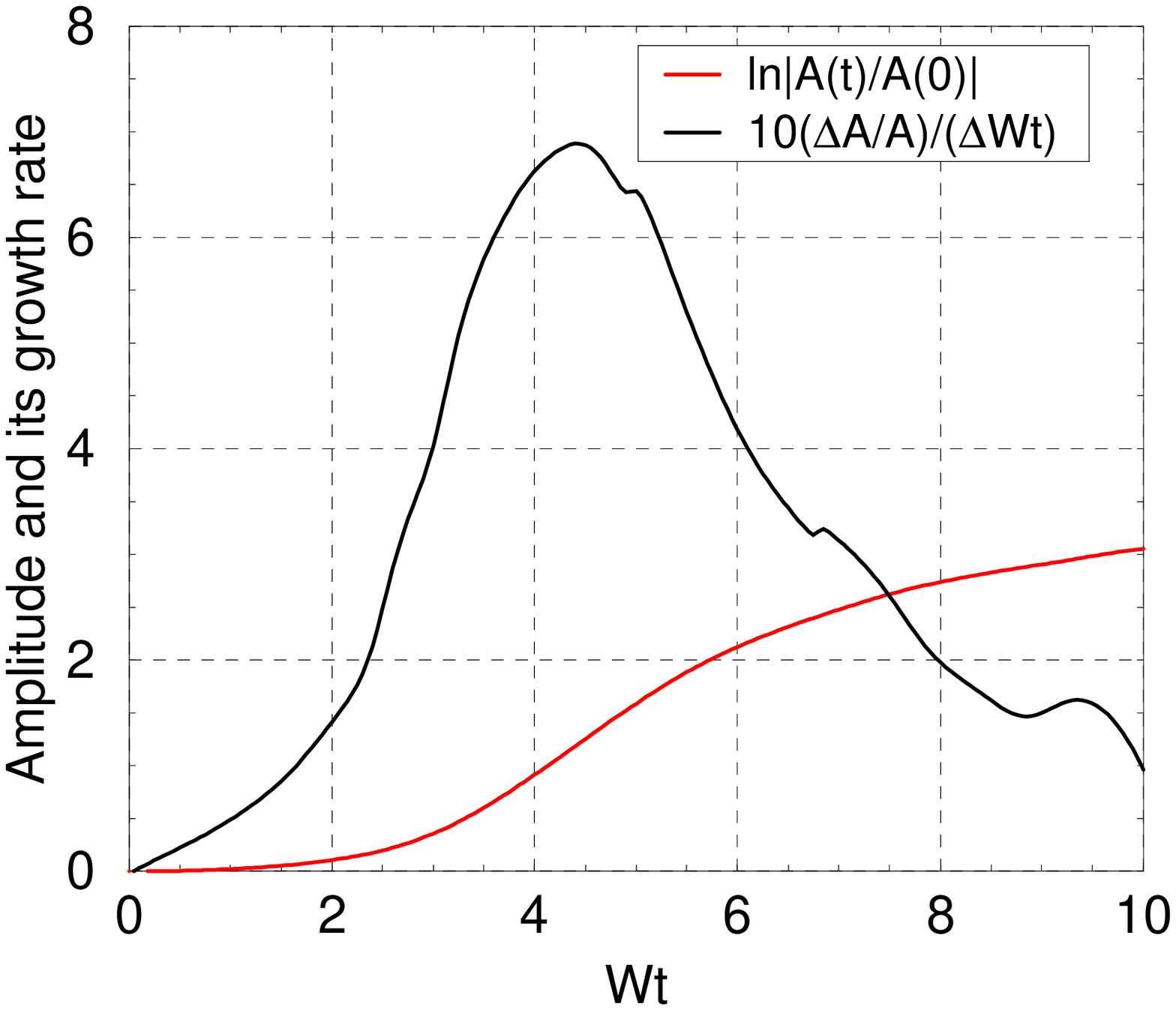}
\end{center}
\end{minipage}
\caption{Instability with nonlinear e-cloud field. Thin beam, one-step wake, 
non-oscillating leading bunch, initial amplitude of other 
bunches $\,A_{Ni}=1\,$, nonlinearity $\,\epsilon=-0.001$.
\\
Left-hand graph: amplitudes of 20 bunches vs time. 
Saturation appears at $\,|\epsilon A^2|\simeq 1$.
\\
Right-hand graph: the amplitude averaged across the batch, and its growth rate.}
\end{figure*}
%
%
\begin{figure*}[t!]
\vspace{10mm}
\hspace{-10mm}
\begin{minipage}[b!]{0.45\linewidth}
\begin{center}
\includegraphics[width=85mm]{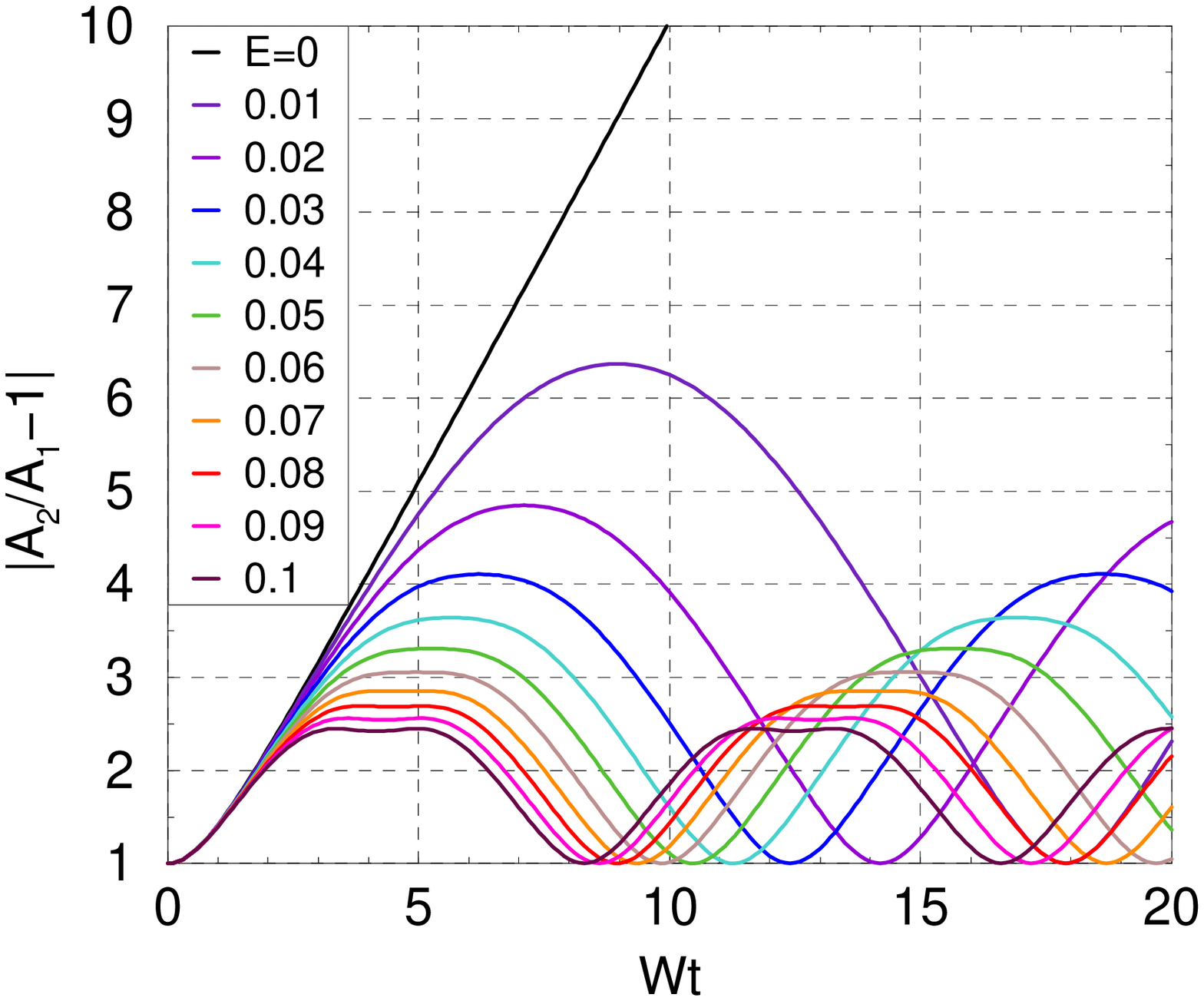}
\end{center}
\end{minipage}
\hspace{0mm}
\begin{minipage}[b!]{0.45\linewidth}
\begin{center}
\includegraphics[width=85mm]{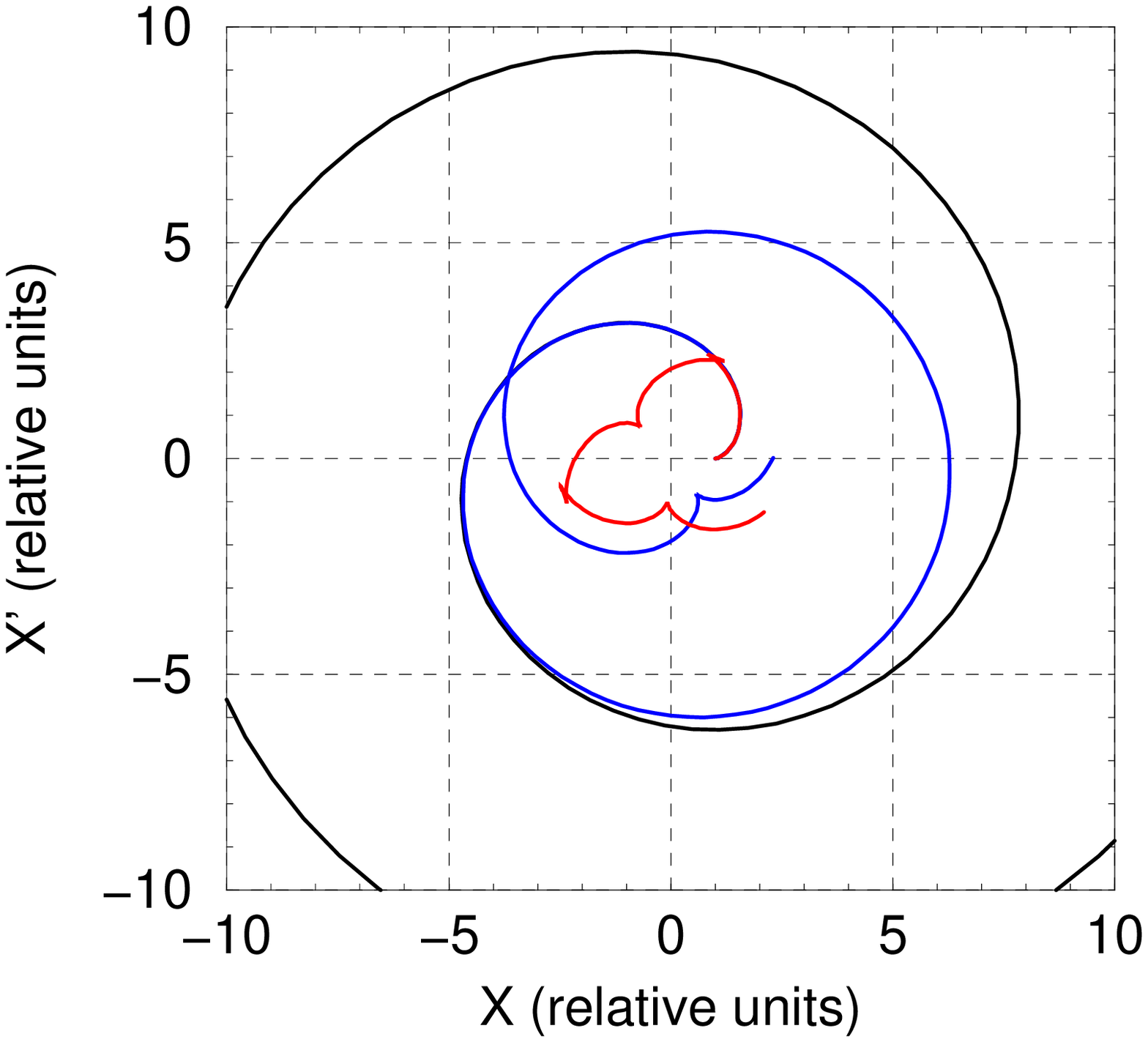}
\end{center}
\end{minipage}
\caption{Oscillations of bunch $\,N=2$. Left-hand: amplitude vs time at different 
values of the nonlinear parameter. Right-hand: phase trajectories of the bunch 
centers  at $\,\epsilon=\,$0~(black),~0.01~(blue),~0.1~(red).
The behavior is typical for nonlinear oscillator excited by external harmonic force.}
\end{figure*}
%
%
\begin{figure*}[t!]
\hspace{-10mm}
\begin{minipage}[b!]{0.45\linewidth}
\begin{center}
\includegraphics[width=85mm]{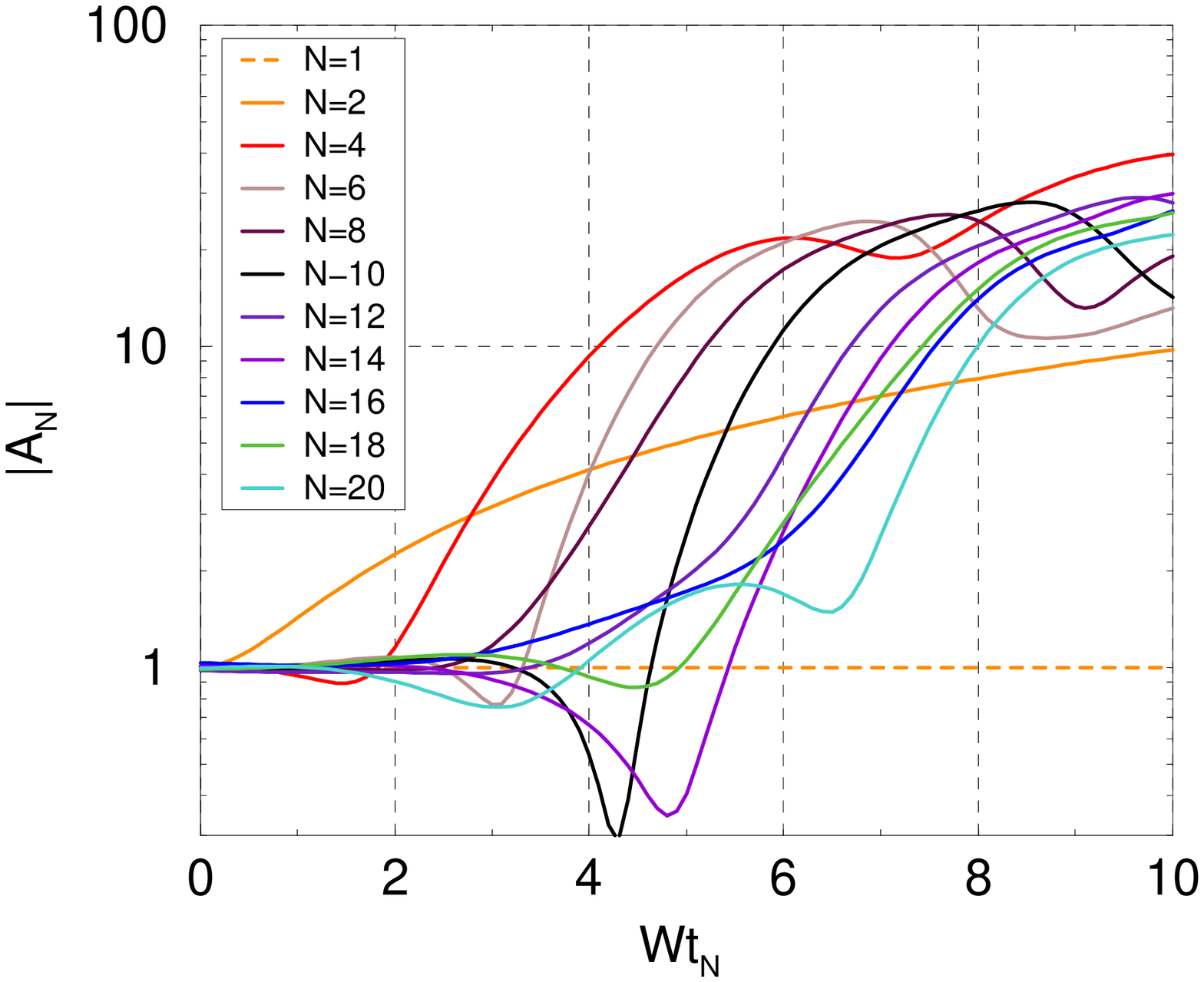}
\end{center}
\end{minipage}
\hspace{0mm}
\begin{minipage}[b!]{0.45\linewidth}
\begin{center}
\includegraphics[width=85mm]{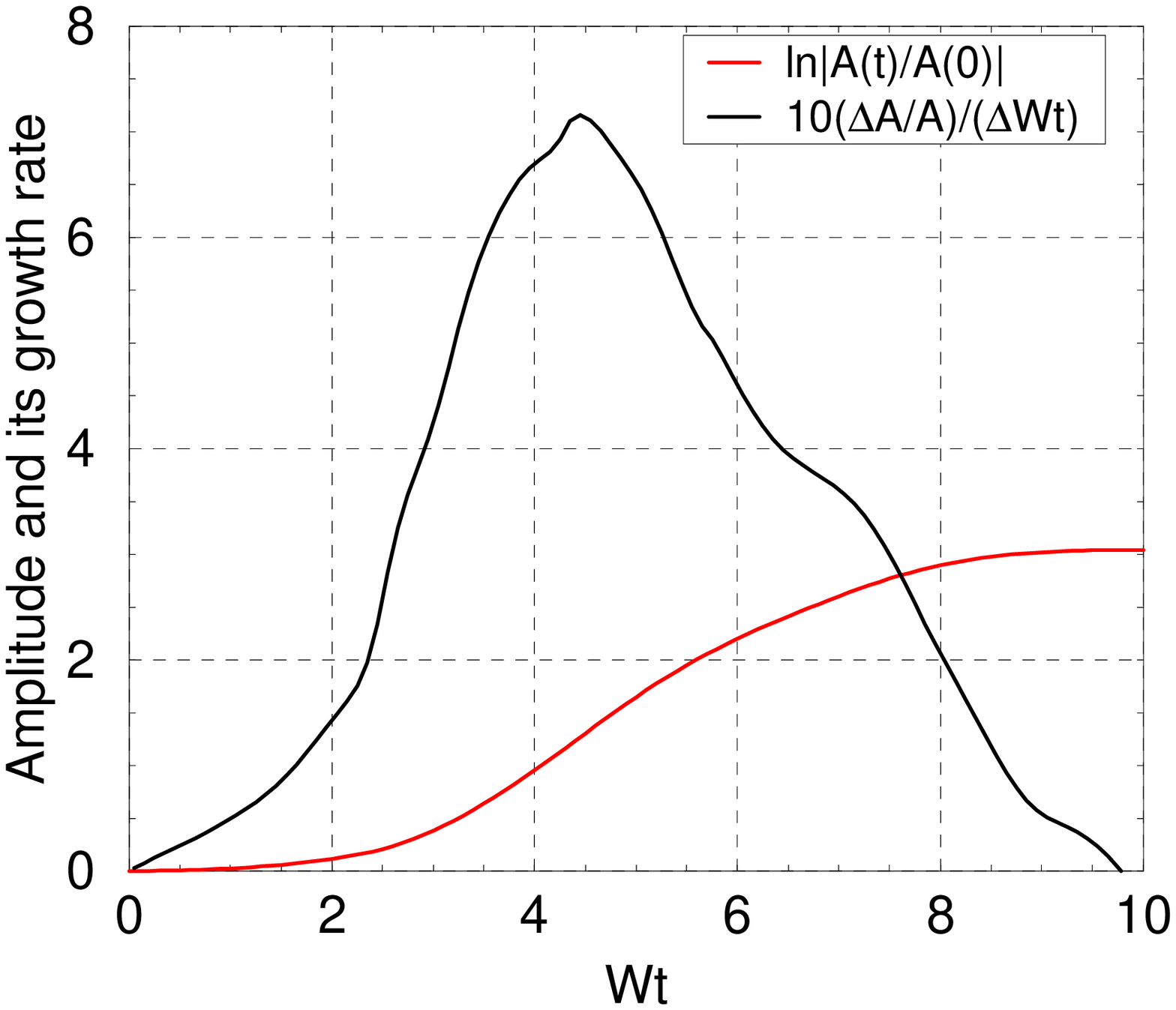}
\end{center}
\end{minipage}
\vspace{5mm}
\caption{The same as in Fig.~8 but the beam is thick: water-bag model of radius 1.
\\
Saturation at $\,|\epsilon A^2|\simeq 1$.}
\end{figure*}
%
%
\begin{figure*}[h!]
\vspace{20mm}
\hspace{-10mm}
\begin{minipage}[b!]{0.45\linewidth}
\begin{center}
\includegraphics[width=85mm]{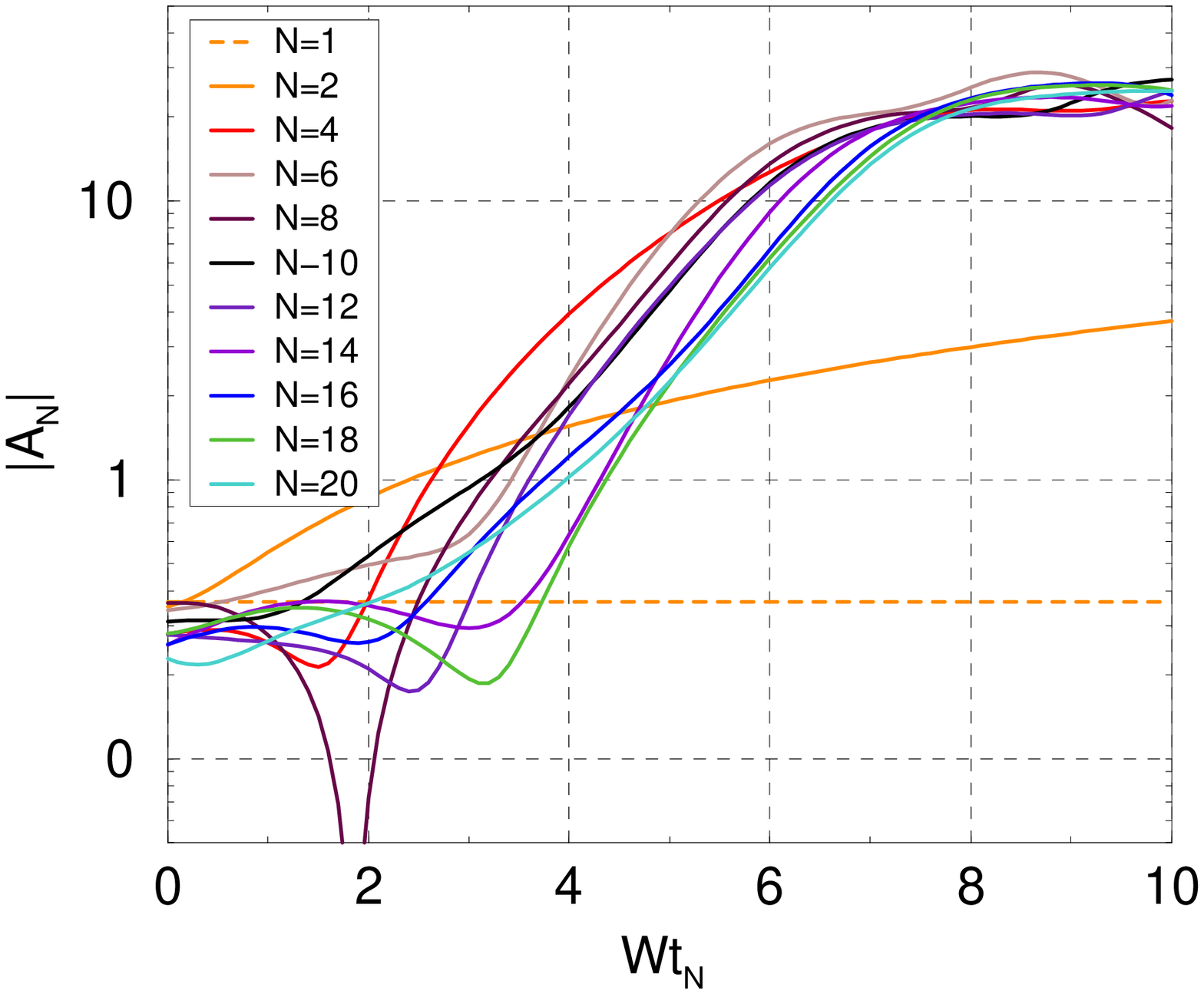}
\end{center}
\end{minipage}
\hspace{0mm}
\begin{minipage}[b!]{0.45\linewidth}
\begin{center}
\includegraphics[width=85mm]{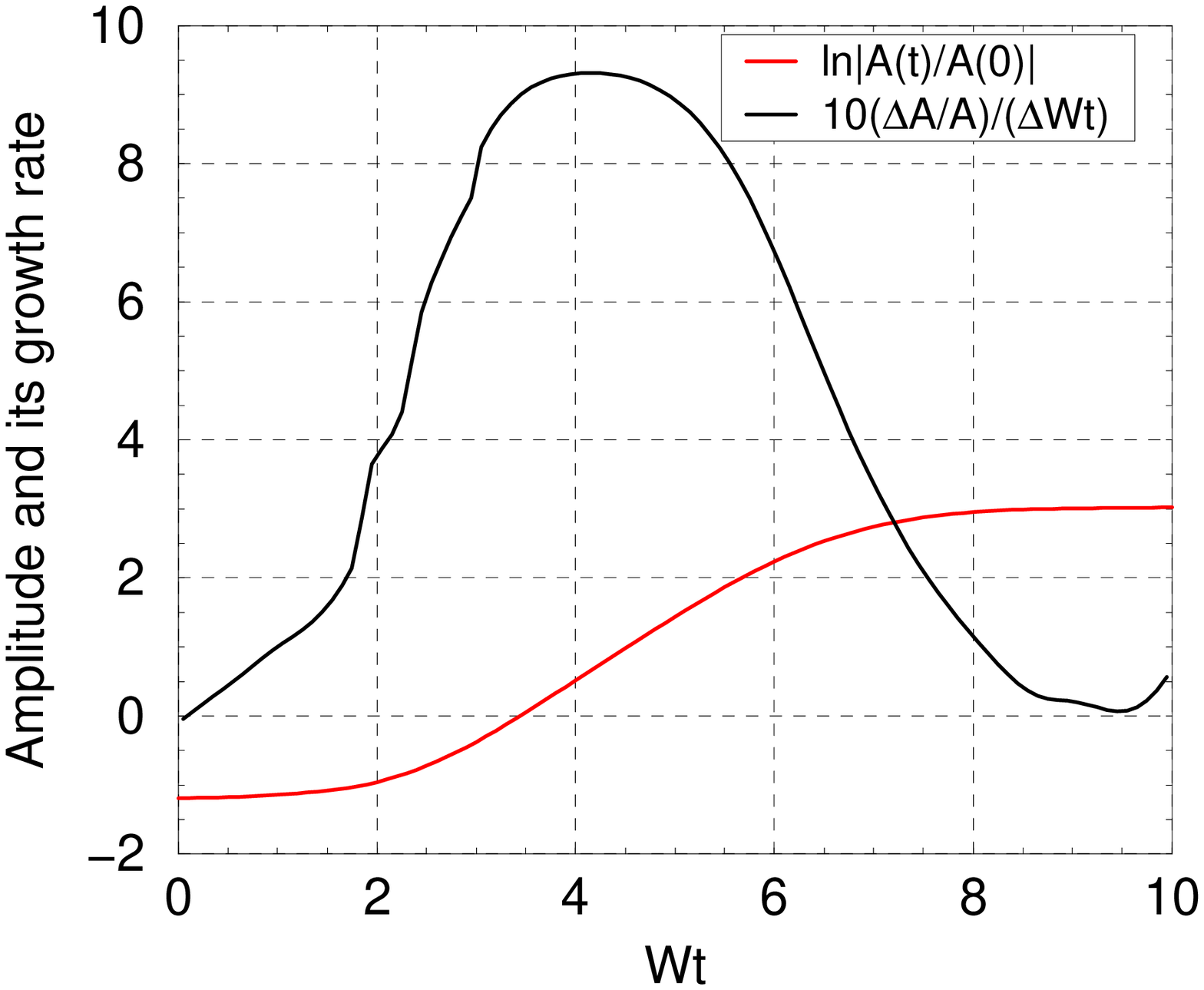}
\end{center}
\end{minipage}
\vspace{5mm}
\caption{The same as in Fig.~10 but the injection errors include a random part:
\\
$A_{Ni}=0.3+0.1\exp(i\phi_N)\,$ with $\,\phi_N\,$ as uniformly distributed random.}
\end{figure*}
%
%
\begin{figure*}[t!]
\hspace{-4mm}
\begin{minipage}[b!]{0.45\linewidth}
\begin{center}
\includegraphics[width=85mm]{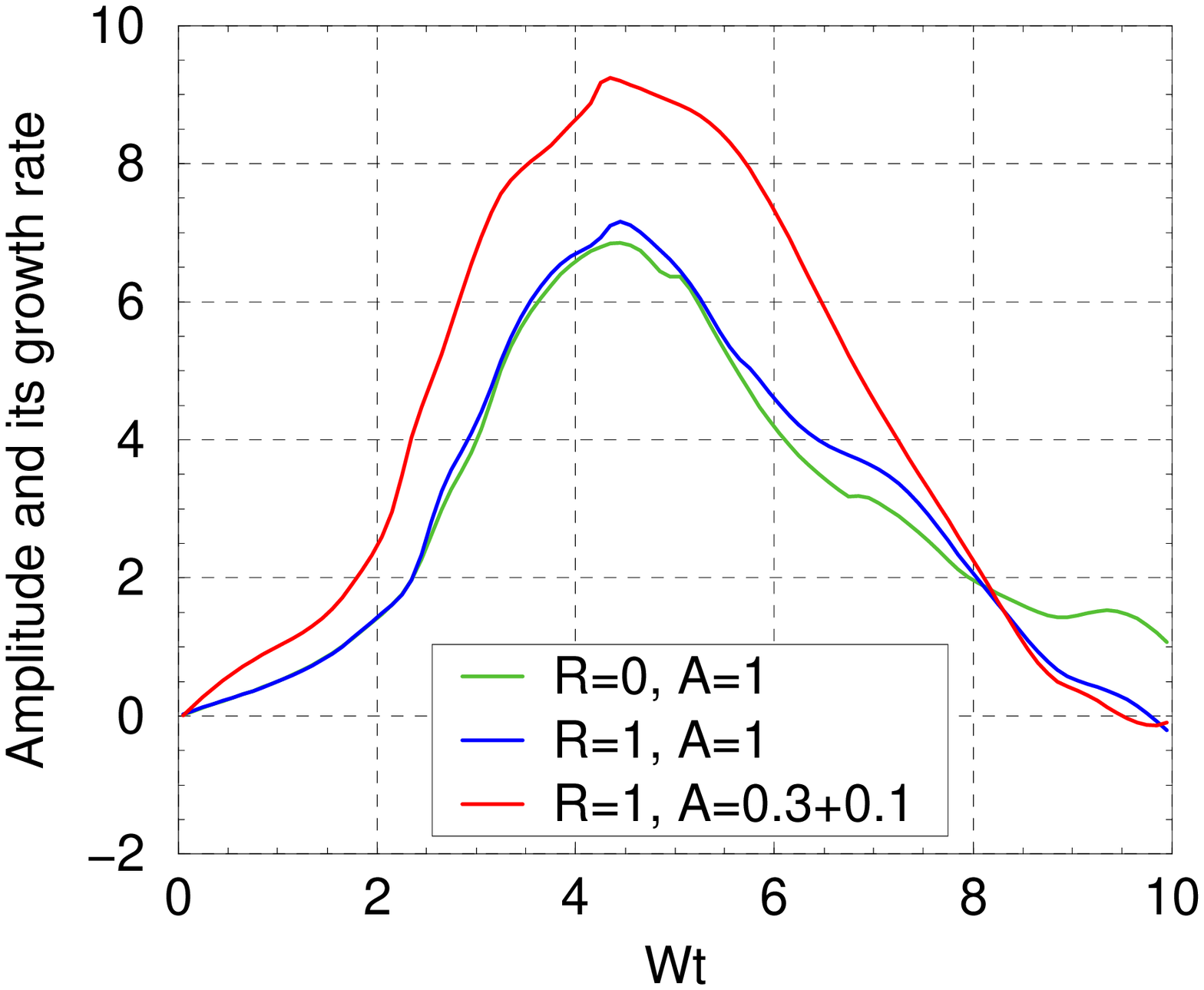}
\end{center}
\end{minipage}
\hspace{0mm}
\begin{minipage}[b!]{0.55\linewidth}
\vspace{12mm}
\includegraphics[width=85mm]{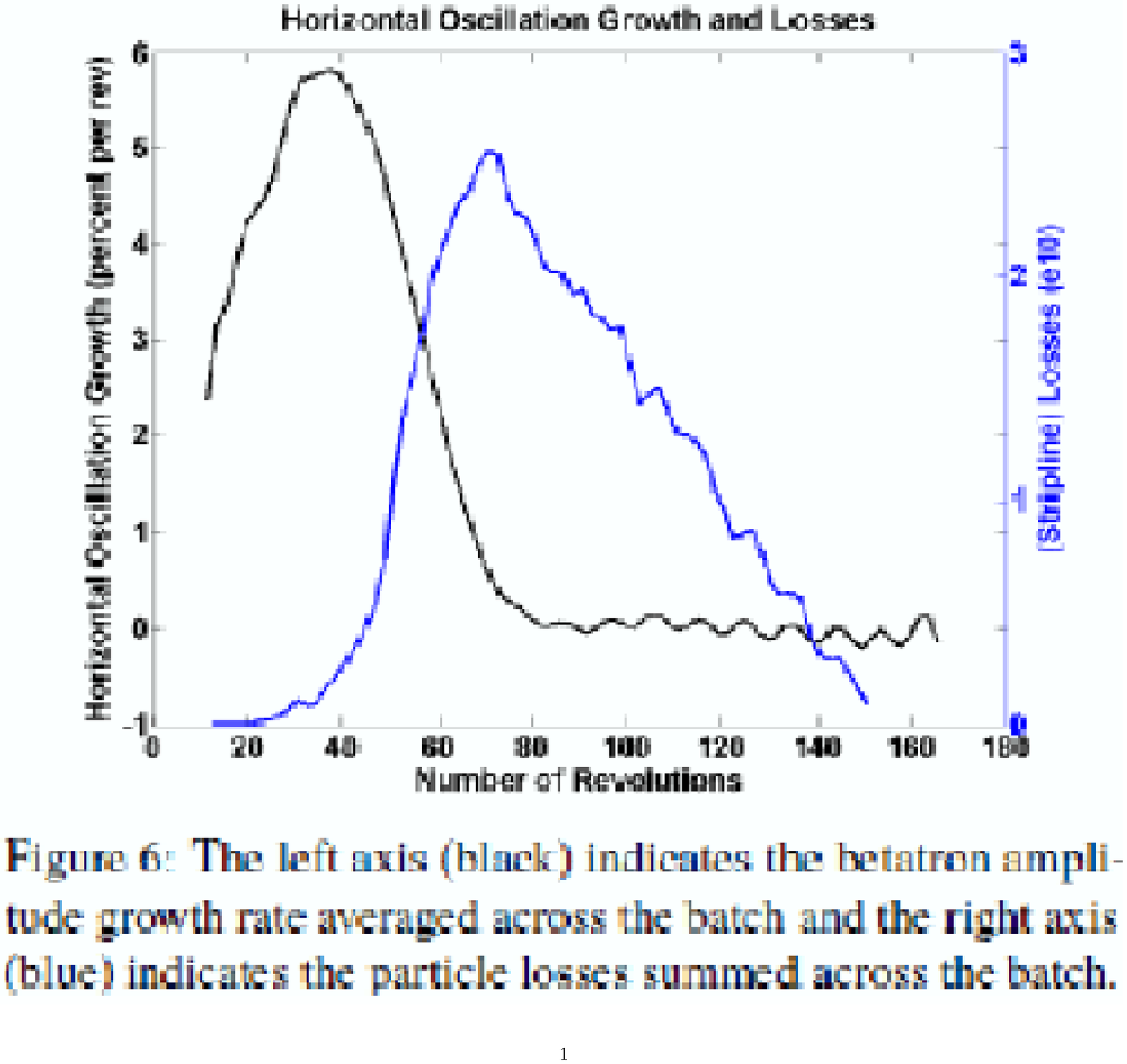}
\end{minipage}
\vspace{0mm}
\caption{Left-hand: the averaged instability growth rate as it has been represented 
in Fig.~8,~10,~11. Right-hand: measured instability growth rate in percent per 
turn \cite{MAIN} }
\vspace{5mm}
\end{figure*}
%

The data are collected in left-hand Fig.~12 where the average betatron amplitude 
growth rate averaged across the batch (20 bunches) are represented.
The curves have much in common with each other having a maximum at $Wt\simeq4$
and decreasing to zero at $Wt\simeq10$. 
Similar experimental curve for the Recycler is shown in the right-hand graph which 
is copied from Ref.~\cite{MAIN}. 
By comparison of the plots, one can get the relation of the parameters:
$\,Wt=10\,$ corresponds to 80 revolutions, that is $\,WT_{\rm rev}\simeq 1/8$.
Because of $WT_{\rm rev}=2\pi\Delta Q$ with the one-step wake, 
$\,\Delta Q\simeq 0.02\,$ in these examples.
On the other hand,
%
\begin{equation}
 \Delta Q=\frac{2\pi r_0\rho_e R^2}{Q\beta^2\gamma}\simeq\frac{\rho_e}{10^{14}{\rm m}^3}
\end{equation}
%
that is the e-cloud density can be estimated as 
$\,\rho_e\simeq 2\times 10^{12}/{\rm m}^3$.    

%

\section{Conclusion}

%

Model of electron cloud is developed in the paper to explain the e-cloud 
instability of bunched proton beam in the Fermilab Recycler \cite{MAIN}. 

By this model, e-cloud is an immobile snake 
which density depends on horizontal coordinate and time.
The cloud is composed of e-streams each of them is generated by some 
proton and is remembering its position.

Interaction of proton beam with the cloud can result in an amplification of injection 
errors in form of coherent bunch oscillation growing up from the batch head to its tail.
Spread of the errors from bunch to bunch is one of the conditions of the instability.

Another condition is an approach of the bunch eigentunes which value is proportional 
to the cloud density.
Therefore the amplitude growth accepts a systematic disposition in the batch tail
where the cloud is saturated.
 
The amplitude growth is restricted by nonlinearity of the e-cloud field.

Results of calculations correlate with the experiment qualitatively and
in order of value.


\vspace{5mm}
\begin{center}
{\bf Appendix: REPRINTED FIGURES \cite{MAIN}}
\end{center}
%
\begin{figure}[h!]
\includegraphics[width=100mm]{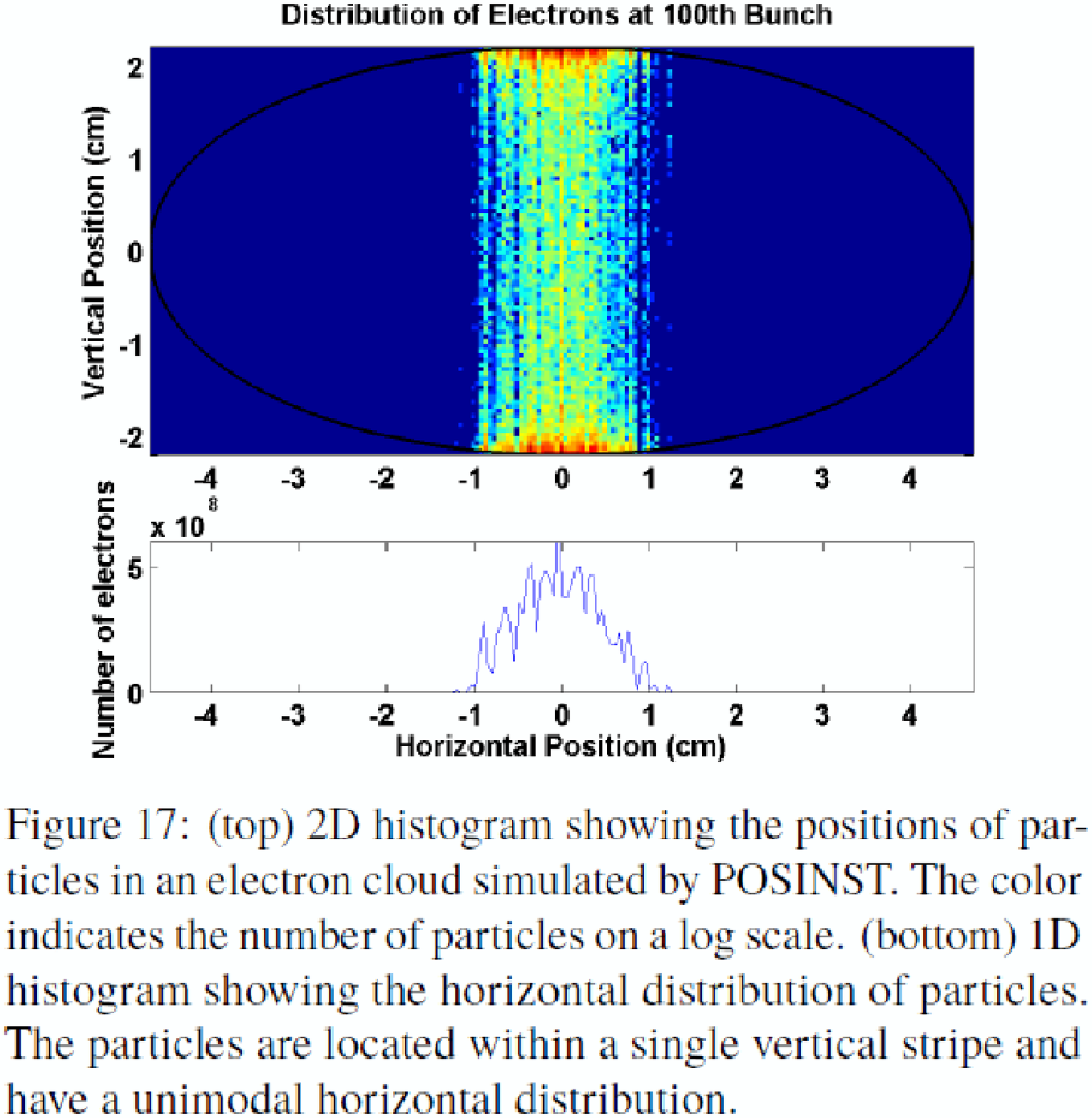}
\end{figure}
%
\begin{center}
{\bf Fig.~A1.~E-cloud transverse cross section simulated by POSINST.}
\end{center}
\newpage
%
\begin{figure}[h!]
\vspace*{0mm}
\includegraphics[width=100mm]{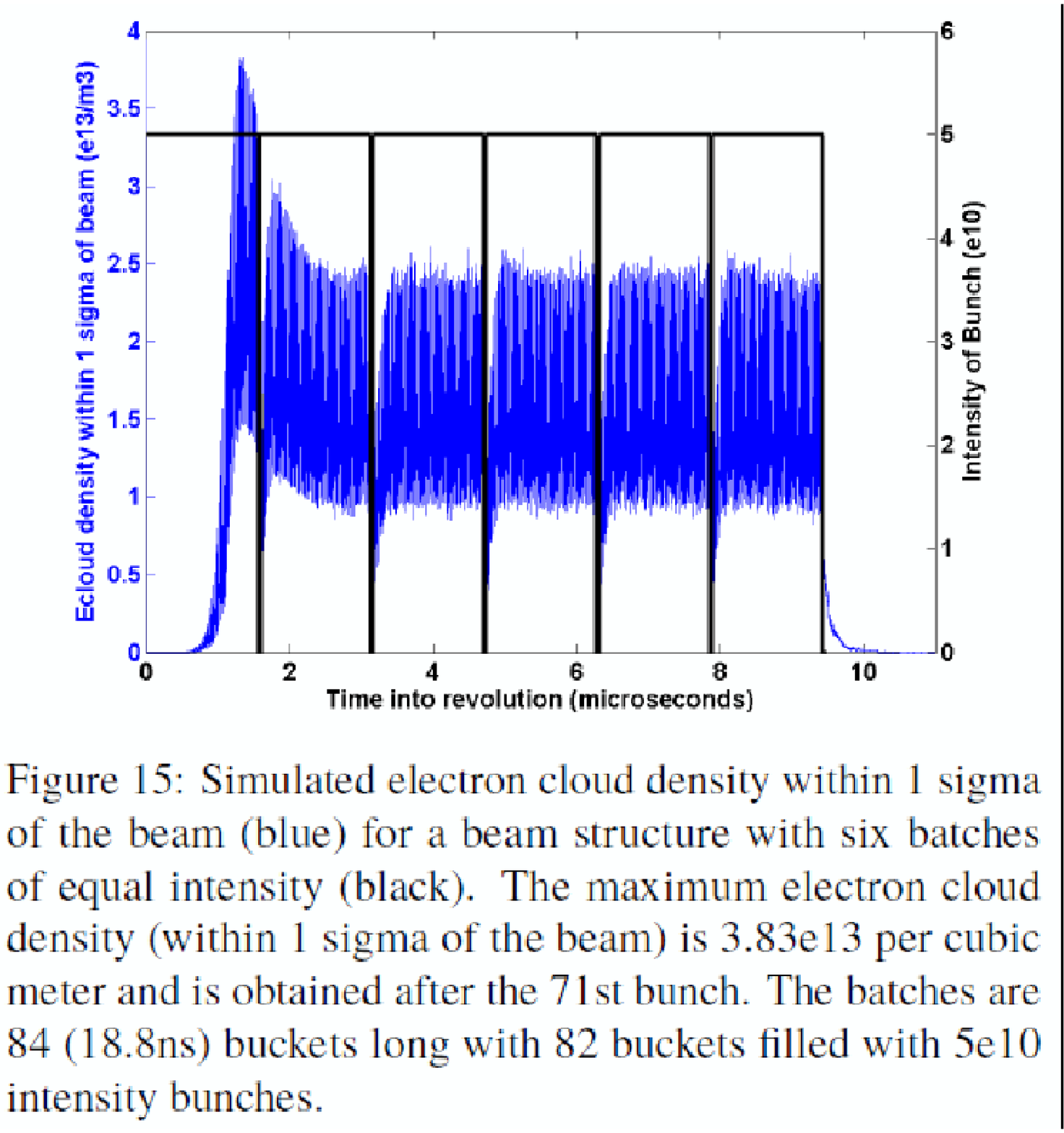}
\vspace{-5mm}
\end{figure}
%
\begin{center}
{\bf Fig.~A2.~E-cloud profile simulated by POSINST}
\end{center}
%
\begin{figure}[h!]
\includegraphics[width=100mm]{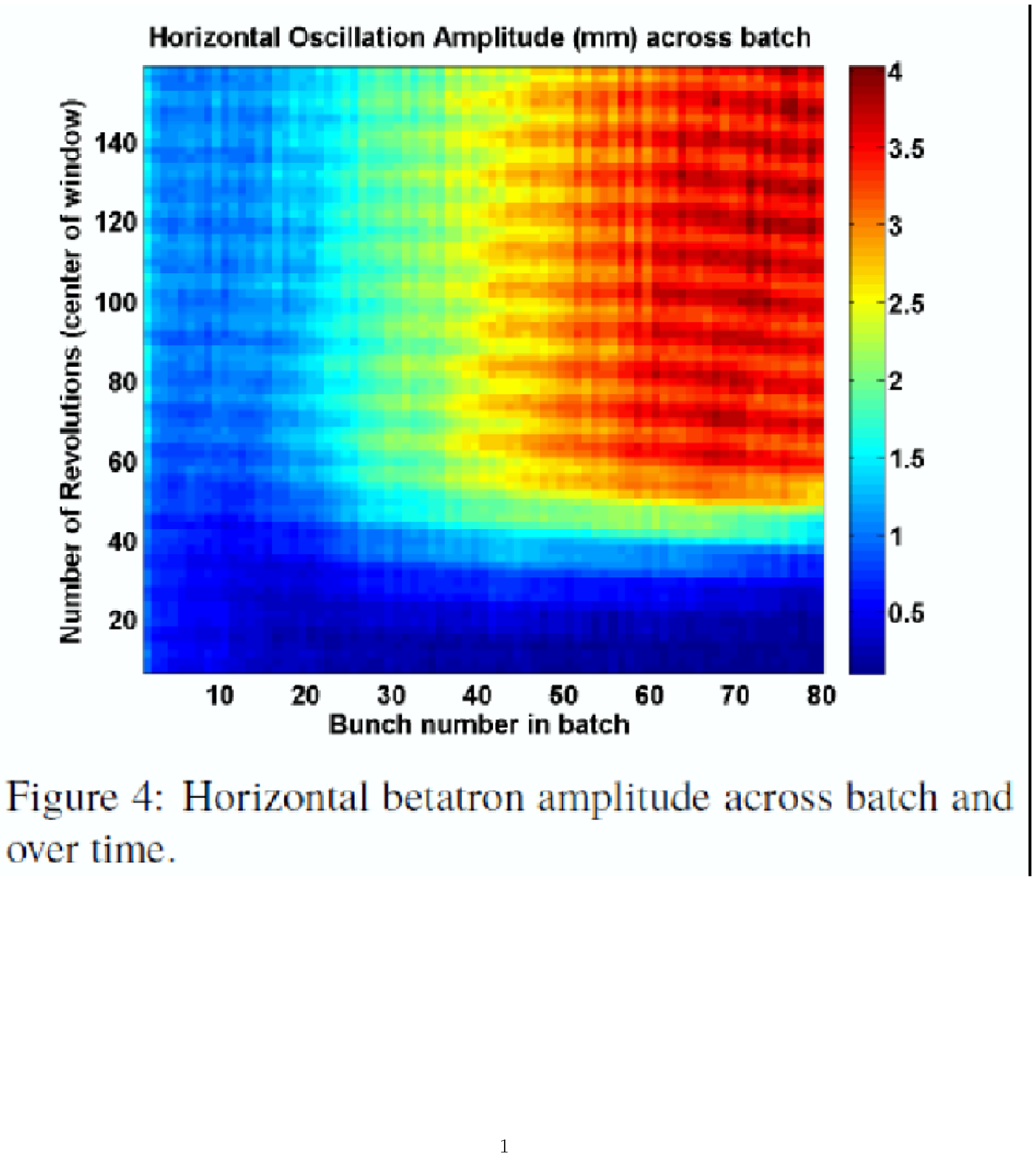}
\vspace{-30mm}
\end{figure}
%
\begin{center}
{\bf Fig.~A3 Horizontal betatron amplitude across batch and over time.}
\end{center} 

\end{document}